# A Note on the Prehistory of Superheavy Elements

Helge Kragh[a]

Centre for Science Studies, Aarhus University, Building 1529, 8000 Aarhus, Denmark

**Abstract**   Artificially produced chemical elements heavier than uranium have been known for more than seventy years and the number of superheavy elements continues to grow. Presently 26 transuranic elements are known. This paper examines the earliest scientific interest in the very heavy elements and the related question of an upper limit of the periodic system. In the period from the 1880s to the early 1930s, three kinds of questions appealed to a minority of physicists, chemists and astronomers: (1) Why is uranium the heaviest known element? (2) Do there exist transuranic or superheavy elements elsewhere in the universe, such as in stellar interiors? (3) Is there a maximum number of elements, corresponding to a theoretical limit for the periodic system? The early attempts to answer or clarify these questions lacked a foundation in nuclear physics, not to mention the total lack of experimental evidence, which explains why most of them were of a speculative nature. Although the speculations led no nothing, they are interesting in their own right and deserve a place in the history of the physical sciences.

## 1  Introduction

How many chemical elements are there? How many can there be? Is there an upper limit to the atomic weight of an element? Over the last couple of decades these questions have been part of several successful large-scale research programmes aiming at synthesizing superheavy elements with an atomic number larger than $Z = 107$. Among the important players in this area of experimental nuclear physics are the Gesellschaft für Schwerionforschung in Darmstadt, Germany, and the Dubna Joint Institute for Nuclear Research near Moscow [Armbruster and Münzenberg 2012; Hofmann 2002]. The present record is Z = 118, provisionally named ununoctium in accordance with the rules of IUPAC, the International Union of Pure and Applied Chemistry. A few atoms of atomic mass number 294 were reported in 2002 and with greater confidence in 2006, by a team consisting of physicists from the Dubna and Lawrence-Livermore laboratories [Oganessian 2006; Sanderson 2006]. The discovery followed a couple of earlier and highly controversial discovery claims that had to be retracted, something which is not uncommon in this area of research [Hofmann 2002, pp. 194-204]. Ununoctium or eka-radon is believed to be real, and the same is the case with element 117, ununseptium or eka-astatine, announced in 2010 [Oganessian 2010]. On the other hand, discovery reports of Z = 122, called unbibium or eka-thorium, remain unconfirmed.

---

[a] e-mail: helge.kragh@ivs.au.dk

   Although the manufacture of superheavy elements by means of heavy-ion collisions is a modern research field, it remains within the older tradition of modern alchemy that started in 1940-1941 with the detection of the first man-made transuranic metals, neptunium (Z = 93) and plutonium (Z = 94). For a long time dominated by Glenn Seaborg and his collaborators at the Lawrence Berkeley National Laboratory (now the Lawrence Livermore National Laboratory), by the early 1960s all elements up to and including the last of the actinides Z = 103 (Lawrencium) had been produced [Weeks 1968, pp. 830-857; Seaborg 1994]. This early tradition grew out of the even earlier first attempts to manufacture transuranic elements, starting with the famous 1934 neutron experiments made by Enrico Fermi, Edoardo Amaldi, Emilio Segré and others. As is well known, although Fermi and his group in Rome did not succeed, for a time they believed to have produced the elements Z = 93 and Z = 94 [Fermi 1934; Sime 2000]. As late as December 1938, in his Nobel Lecture in Stockholm, Fermi referred confidently to the two elements, which he and his collaborators in Rome named "ausonium" and "hesperium" [Fermi 1965, p. 417]. In spite of the failure, the experiments in the 1930s may reasonably be seen as the beginning of experimental research in transuranic or superheavy elements.

It may be less well known that the pioneering experiments of Fermi, Seaborg and others did not mark the beginning of scientific interest in elements heavier than uranium. The purpose of this note is to describe some even older ideas of transuranic elements, including suggestions concerning the maximum number of chemical elements. Such ideas can be found more than a century ago, before quantum theory and the nuclear atom, if most often in the form of speculations rather than scientifically informed predictions [Van Spronsen 1969, pp. 329-337; Kragh and Carazza 1995]. The scattered proposals were either speculative – and some of them *very* speculative – or based on theories that have long ago become obsolete. Nonetheless, they belong to the history of science no less than the later and more scientific ideas out of which the modern research in superheavy elements has grown.

The "prehistoric" interest in the upper limit of the periodic system and elements heavier than uranium may be divided in three classes. First, there were chemical claims and speculations before the introduction of the nuclear atom. Second, the Bohr-Sommerfeld theory of atomic structure led to suggestions of a highest atomic number. And third, there were in the 1920s several speculations about superheavy elements in the stars and the nebulae. Generally, the subject under investigation invites an interdisciplinary approach. It is only by looking at the history of both physics, chemistry and astronomy that one can obtain a reasonably full picture of how scientists in the pre-1940 period considered the question of superheavy elements.

## 2 Before the nuclear atom

The periodic system of the elements introduced by Dmitrii I. Mendeleev and Lothar Meyer in 1869 soon resulted in speculations about the cause of the system, the number and locations of the elements in it, and the possible existence of elements lighter than hydrogen and heavier than uranium. The question concerning subhydrogenic elements has its own fascinating history, but I shall keep to the other limit of the periodic system. In the period from about 1880 to 1915 there were numerous attempts to extend or "explain" Mendeleev's table, and some of these speculative attempts included "predictions" of a highest atomic weight.[1] In this period, before the concept of the atomic number had been introduced and accepted, the atomic weight was universally considered the defining parameter of a chemical element.

A few of the early chemists believed to have found a mathematical or numerological reason for the upper limit of the system of the elements, such as given by $A \cong 240$ corresponding to uranium. Discovered in 1801 by the German chemist Martin Heinrich Klaproth, uranium had been known for decades to be the heaviest of the elements. Why? It could be that there was some theoretical reason for it, or it could be that even heavier elements existed but had escaped detection. According to Edmund Mills, a Glasgow professor of technical chemistry, the atomic weights of all the known elements except hydrogen could be represented with "extremely close agreement" by the formula

$$y = 15(p - 0.9375^x) \, ,$$

---

[1] For more information about the early interest in transuranic elements, whether based on experiments or theoretical speculations, see [Quill 1938], [Tsaletka and Lapitskii 1960], and [Karpenko 1980]. None of these reviews mention the role that superheavy elements played in astrophysical theories.

where *x* is an integer and *p* a group number ranging between 1 and 16. For the group to which uranium belongs, $p = 16$. This piece of numerology led him to suggest that it is "easy to conceive the existence of an upper limit to our existing system." For *x* tending towards infinity, the result becomes 240, which is indeed in close agreement with the experimental value A = 239.70 known at the time. "Hence 240 can hardly fail to be a critical number in, and may very probably be the upper limit of, our existing system," he commented [Mills 1884, p. 399; Mills 1886]. Of course, Mills' reasoning was devoid of empirical content. Any number can be represented by the formula if only suitable values for *p* and *x* are chosen, and so it is no wonder that also the atomic weight of uranium turns up.

The approach of the recognized German chemist Victor Meyer, professor at Göttingen University, was hardly more scientific than Mills'. In an address of 1889 he noted "the peculiar coincidence" that Mendeleev's table indicated two small periods of seven elements each and five larger ones of seventeen elements. To these should be added hydrogen, and thus the number of possible elements came out as $2 \times 7 + 5 \times 17 + 1 = 100$. "As far as positive data are at hand," said the German professor, "they indicate exactly the number mentioned [100] and nothing points beyond it" [Meyer 1889, p. 112]. The kind of dubious reasoning exemplified by Mills and Meyer was followed only by a few chemists. One of them was Sima Losanitsch, a professor of organic chemistry in Belgrade, who in 1906 published a booklet in which he not only proposed several elements heavier than uranium but also elementary particles much lighter than the hydrogen atom [Losanitsch 1906]. However, the large majority of chemists refrained from speculating about the limits of the periodic system or proposing transuranic elements from theoretical reasons. They were aware of the questions, but without considering them very important. As the British chemist William A. Tilden pointed out in 1910, "there is nothing in theory to preclude the expectation of additions of new substances to either extremity of the series," that is, elements lighter than hydrogen or heavier than uranium [Tilden 1910, pp. 56-57]. On the other hand, he found the existence of such elements unlikely. Radioactivity indicated that the heaviest atoms were unstable and that the limit was at uranium at an atomic weight about 240. Although one might imagine still heavier atoms decaying to uranium, there was no reason to do so. There is little doubt that Tilden's view was broadly accepted among both chemists and physicists.

The discovery of radioactivity in 1896 stimulated chemists to reexamine the confusing properties of the heavy elements, with the result that a few of them thought to have discovered new elements with atomic weight greater than uranium's. The prominent Czech chemist Bohuslav Brauner, a friend of Mendeleev and an expert in the chemistry of the rare earth metals, believed that thorium was a complex substance. In experiments with thorium salts he found in 1901 a small fraction of atomic weight $A = 280.7$ as compared to the main fraction's $A = 234.6$ [Brauner 1901]. Although he concluded that thorium was a complex substance, he did not explicitly propose the $A = 280.7$ fraction as a new element heavier than uranium.

Across the Atlantic, Charles Baskerville at the University of North Carolina made experiments of a similar nature, reaching the same conclusion. In 1904 he suggested that the heavy fraction, the atomic weight of which he determined to 255.6, was a new quadrivalent element for which he proposed the name "carolinium" [Baskerville 1904; Brauner and Baskerville 1904]. Although Baskerville was convinced that he had discovered a transuranic element, he realized that it lacked confirmation in the form of spectral analysis and he made no attempt to place carolinium in the periodic table.[2] Brauner considered the element a result of American sensationalism. At any rate, carolinium was but a brief parenthesis in the history of chemistry. It failed to win recognition in the chemical community and suffered the same fate as helvetium, oceanium, austrium, coronium and numerous other spurious elements: a name without a reality [Karpenko 1980]. Carolinium is worth mention only because it may have been the first empirical claim of an element heavier than uranium.

---

[2] Losanitch [1906] placed carolinium in one of his periodic tables, assigning it atomic weight 254 and symbol Cn.

The method of X-ray spectroscopy based on the atomic number made it possible to identify elements more precisely and in smaller amounts than previously. The British chemists Frederick H. Loring and Gerald J. F. Druce were the first to use the method in searching for $Z = 93$, and in a series of papers in *Chemical News* of 1925 they suggested to have detected in manganese minerals two spectral lines originating from the element [Loring 1926]. They wisely avoided claiming the evidence conclusive or proposing a name for the new element. Their two lines proved to belong to other elements. Nine years later the Czech chemical engineer Odolen Koblic announced to have discovered a transuranic element in the uranium mineral pitchblende using traditional chemical fractionation methods. He concluded that the element was $Z = 93$, that it had an atomic weight about 240, and that it was a higher homologue of rhenium. "All examinations carried out bore witness to my successful achievement in isolating the supposed element no. 93, which I name bohemium (Bo) in honour of my fatherland" [Karpenko 1980, p. 89]. Koblic's bohemium was as short-lived as Baskerville's carolinium. X-ray examinations made by Ida Noddack in Berlin failed to detect any lines indicating a new element. Within a month after its announcement, Koblic admitted his error and withdrew his claim [Speter 1934]. Noddack not only killed Koblic's element 93, she also and more importantly objected to Fermi's suggestion of having produced the element.[3]

---

[3] Noddack's paper "On element 93" in the September 1934 issue of *Zeitschrift für angewandte Chemie* is translated in [Graetzer and Anderson 1971, pp. 16-18]. Apart from demonstrating that Fermi's interpretation of the neutron experiments was untenable, she

## 3. Quantum-based suggestions

The years 1911-1913 constituted a quiet revolution in the conception of chemical elements, a result of Ernest Rutherford's nuclear model, the recognition of isotopy, Henry Moseley's determinations of X-ray spectra, and Niels Bohr's quantum theory of atomic structure. With the introduction of the atomic number Z, corresponding to the positive charge of the atomic nucleus, followed a new definition of an element in better agreement with the periodic system. While elements lighter than hydrogen made sense according to the older definition, they were now ruled out. On the other hand, the replacement of the atomic weight with the atomic number did not change the situation with regard to possible transuranic elements: they might exist, or they might not exist.

The Bohr atom offered a more realistic picture of the atom than the earlier Thomson atom and made it possible, for the first time, to compare atomic models with the actual properties of the elements. It also made it possible to come up with scientifically based answers, rather than mere speculations, to the question of an upper limit to the periodic table. In Bohr's revised atomic theory of 1921-1923 the orbit of an electron in an atom was characterized by two quantum numbers, the principal quantum number $n$ and the azimuthal quantum number $k$ [Kragh 2012, pp. 271-302]. He designated the state as $n_k$ and for $x$ electrons moving in the same orbital state he used the notation $(n_k)^x$. For example, the lithium atom in its ground state would be $(1_1)^2(2_1)^1$. Bohr suggested electron configurations for all the elements in the periodic system, even the heaviest ones. He predicted a second rare-earth series analogous to the lanthanides, but without being able to determine

---

also suggested as an alternative interpretation that the uranium nucleus might have broken up in two or more fractions. Her paper was ignored by both physicists and radiochemists, and Noddack was only rehabilitated as a precursor of the fission hypothesis after her death in 1978.

its beginning. In his version of the periodic system he placed the new series beyond uranium rather than placing it as an actinide series including uranium.

| | $1_1$ | $2_1 2_2$ | $3_1 3_2 3_3$ | $4_1 4_2 4_3 4_4$ | $5_1 5_2 5_3 5_4 5_5$ | $6_1 6_2 6_3 6_4 6_5 6_6$ | $7_1 7_2$ |
|---|---|---|---|---|---|---|---|
| 1 H | 1 | | | | | | |
| 2 He | 2 | | | | | | |
| 3 Li | 2 | 1 | | | | | |
| 4 Be | 2 | 2 | | | | | |
| 5 B | 2 | 2(1) | | | | | |
| — — | — | — — | | | | | |
| 10 Ne | 2 | 4 4 | | | | | |
| 11 Na | 2 | 4 4 | 1 | | | | |
| 12 Mg | 2 | 4 4 | 2 | | | | |
| 13 Al | 2 | 4 4 | 2 1 | | | | |
| — — | — | — — | — — | | | | |
| 18 A | 2 | 4 4 | 4 4 | | | | |
| 19 K | 2 | 4 4 | 4 4 | 1 | | | |
| 20 Ca | 2 | 4 4 | 4 4 | 2 | | | |
| 21 Sc | 2 | 4 4 | 4 4 1 | (2) | | | |
| 22 Ti | 2 | 4 4 | 4 4 2 | (2) | | | |
| — — | — | — — | — — — | — | | | |
| 29 Cu | 2 | 4 4 | 6 6 6 | 1 | | | |
| 30 Zn | 2 | 4 4 | 6 6 6 | 2 | | | |
| 31 Ga | 2 | 4 4 | 6 6 6 | 2 1 | | | |
| — — | | — — | — — — | — — | | | |
| 36 Kr | 2 | 4 4 | 6 6 6 | 4 4 | | | |
| 37 Rb | 2 | 4 4 | 6 6 6 | 4 4 | 1 | | |
| 38 Sr | 2 | 4 4 | 6 6 6 | 4 4 | 2 | | |
| 39 Y | 2 | 4 4 | 6 6 6 | 4 4 1 | (2) | | |
| 40 Zr | 2 | 4 4 | 6 6 6 | 4 4 2 | (2) | | |
| — — | — | — — | — — — | — — — | — | | |
| 47 Ag | 2 | 4 4 | 6 6 6 | 6 6 6 | 1 | | |
| 48 Cd | 2 | 4 4 | 6 6 6 | 6 6 6 | 2 | | |
| 49 In | 2 | 4 4 | 6 6 6 | 6 6 6 | 2 1 | | |
| — — | | — — | — — — | — — — | — — | | |
| 54 X | 2 | 4 4 | 6 6 6 | 6 6 6 | 4 4 | | |
| 55 Cs | 2 | 4 4 | 6 6 6 | 6 6 6 | 4 4 | 1 | |
| 56 Ba | 2 | 4 4 | 6 6 6 | 6 6 6 | 4 4 | 2 | |
| 57 La | 2 | 4 4 | 6 6 6 | 6 6 6 | 4 4 1 | (2) | |
| 58 Ce | 2 | 4 4 | 6 6 6 | 6 6 6 1 | 4 4 1 | (2) | |
| 59 Pr | 2 | 4 4 | 6 6 6 | 6 6 6 2 | 4 4 1 | (2) | |
| — — | — | — — | — — — | — — — — | — — — | — | |
| 71 Cp | 2 | 4 4 | 6 6 6 | 8 8 8 8 | 4 4 1 | (2) | |
| 72 — | 2 | 4 4 | 6 6 6 | 8 8 8 8 | 4 4 2 | (2) | |
| — — | — | — — | — — — | — — — — | — — — | — | |
| 79 Au | 2 | 4 4 | 6 6 6 | 8 8 8 8 | 6 6 6 | 1 | |
| 80 Hg | 2 | 4 4 | 6 6 6 | 8 8 8 8 | 6 6 6 | 2 | |
| 81 Tl | 2 | 4 4 | 6 6 6 | 8 8 8 8 | 6 6 6 | 2 1 | |
| — — | | — — | — — — | — — — — | — — — | — — | |
| 86 Em | 2 | 4 4 | 6 6 6 | 8 8 8 8 | 6 6 6 | 4 4 | |
| 87 — | 2 | 4 4 | 6 6 6 | 8 8 8 8 | 6 6 6 | 4 4 | 1 |
| 88 Ra | 2 | 4 4 | 6 6 6 | 8 8 8 8 | 6 6 6 | 4 4 | 2 |
| 89 Ac | 2 | 4 4 | 6 6 6 | 8 8 8 8 | 6 6 6 | 4 4 1 | (2) |
| 90 Th | 2 | 4 4 | 6 6 6 | 8 8 8 8 | 6 6 6 | 4 4 2 | (2) |
| — — | — | — — | — — — | — — — — | — — — | — — — | — |
| 118 ? | 2 | 4 4 | 6 6 6 | 8 8 8 8 | 8 8 8 8 | 6 6 6 | 4 4 |

Bohr's atomic structures of 1922, including Z = 118.

On a few occasions Bohr went further, into the *terra incognita* of transuranium elements. Thus, in a series of lectures in Göttingen in June 1922 he wrote down the electron configuration of uranium, and then announced to his audience: "We might proceed further … and construct hundreds or thousands of elements." Perhaps feeling that his enthusiasm had carried him away from his usual soberness, he added, "however, that is not the task of physics, which deals only with things that can be put to experimental test" [Rud Nielsen 1977, p. 405]. Nonetheless, he did not hesitate to predict the configuration of the hypothetical element Z = 118, stating that it would be a noble gas with chemical properties similar to radon. His suggestion was this:

$$(1_1)^2 \cdot (2_1)^4(2_2)^4 \cdot (3_1)^6 \cdot (3_2)^6(3_3)^6 \cdot (4_1)^8(4_2)^8(4_3)^8(4_4)^8 \cdot (5_1)^8(5_2)^8(5_3)^8(5_4)^8 \cdot (6_1)^6 \cdot (6_2)^6(6_3)^6 \cdot (7_1)^4(7_2)^4$$

Also in his Nobel Lecture later the same year Bohr included the hypothetical element Z = 118 in his table with the configurations of the elements, but without commenting on its properties [Bohr 1923a]. A few years later he asked Yoshio Nishina, a physicist from Japan staying at Bohr's institute in Copenhagen, to examine by means of X-ray spectroscopy whether there might be, as he suspected, elements of Z = 93, 94 or 96 homologous to uranium [Kim 2007, p. 26]. It is unknown if Nishina looked for these elements in uranium minerals. If he did, nothing came out of it.

Written as the number of electrons in the various "shells" or energy levels $n$ from 1 to 7, the configuration for Z = 118 derived by Bohr in 1922 was

2, 8, 18, 32, 32, 18, 8

It is interesting to observe that the very same structure was found by Clinton Nash of the University of New England when he, more than eighty years later, calculated the electronic structure of ununoctium [Nash 2005]. However, contrary to the expectation of Bohr, Nash's calculations indicated that element 118 was far more active than radon and probably not a gas under normal conditions.

Bohr probably did not believe in the existence of superheavy elements. He subscribed to the generally accepted view that "nuclei of atoms with a total charge greater than 92 will not be sufficiently stable to exist under conditions where the elements can be observed" [Bohr 1924, p. 112]. Still, in the early 1920s the possibility of transuranium elements and the question of an upper limit of the periodic system were subjects discussed in his institute. One indication is a note of 1923 written by the young Norwegian physicist Svein Rosseland, who stayed at Bohr's institute 1920-1924. Rosseland, who would later become a leader of astrophysics, investigated the hypothesis that radioactivity is caused by the influence of the orbital electrons. According to the Bohr-Sommerfeld atomic theory the shortest distance between a nucleus and an elliptically moving electron would be attained for electrons with $k = 1$ and be approximately given by

$$r = \frac{a_0}{2Z}(1 - \alpha^2 Z^2),$$

where $a_0$ is the Bohr radius and $\alpha$ the fine-structure constant given by

$$a_0 = \frac{h^2}{4\pi^2 m e^2} \quad \text{and} \quad \alpha = \frac{2\pi e^2}{hc}$$

Since $r$ will diminish with increasing Z, and the size of the nucleus will increase, Rosseland suggested that there would exist an upper limit for the atomic number. Although he did not calculate this limit, he found it unlikely that there would exist elements with atomic numbers much larger than 92, for in this case "the electrons in question would have to collide with the nucleus" [Rosseland 1923].

Rosseland's speculations were undoubtedly cleared with Bohr, who later the same year stated without proof that an electron in a $n_k$ orbit would fall into the nucleus if

$$\frac{Z}{k} \geq \frac{hc}{2\pi e^2} = \frac{1}{\alpha} \cong 137$$

For $k = 1$, this means that $Z < 137$. Bohr commented that "the electron in these [heavy] elements comes to distances from the nucleus of the same order of magnitude as the value of the nuclear dimensions … [and] this circumstance alone offers a hint toward an understanding of the limitation in the atomic number of existing elements" [Bohr 1923b, p. 266]. Bohr's remark was elaborated upon by Sommerfeld in the fourth edition of his classical work *Atombau und Spektrallinien*, using the relativistic energy expression he had derived in his fine-structure theory for one-electron atoms [Sommerfeld 1924, pp. 465-468]. With the radial quantum number given by $n_r = n - k$, Sommerfeld expressed the energy as

$$1 + \frac{E}{mc^2} = \left\{1 + \frac{\alpha^2 Z^2}{\left[n_r + \sqrt{k^2 - \alpha^2 Z^2}\right]^2}\right\}^{-1/2}$$

For a circular orbit ($n = k$, or $n_r = 0$) this gives

$$1 + \frac{E}{mc^2} = \sqrt{1 - \alpha^2 (Z/k)^2}$$

In order that the energy be real, one must then have

$$1 - \alpha^2 \left(\frac{Z}{k}\right)^2 \geq 0,$$

which corresponds to Bohr's condition and implies $Z \leq 137$. For $k > \alpha Z$, electrons of momentum $p$ move in rotating elliptic orbits with a perihelion motion given by

$$\gamma^2 = 1 - \frac{1}{p^2}\frac{Ze^2}{c^2} = 1 - \left(\frac{k}{\alpha Z}\right)^2$$

In 1924 Sommerfeld proved that if $k < \alpha Z$ the motion would not be elliptical, but the electron would instead perform a spiral motion around the nucleus, approaching it almost with the speed of light. For $k = 1$, $Z = 137$ would therefore be the limit between allowed elliptical orbits and forbidden spiraling orbits.

In the early 1920s there was much discussion about "half-quanta" or half-integral quantum numbers such as suggested by data from molecular spectroscopy and the anomalous Zeeman effect. Bohr denied that the azimuthal quantum number could be $k = ½$ or attain other half-integral values, which he thought contradicted his correspondence principle. He found support for his view in the heavy elements, the reason being that a K-electron with $k = ½$ would imply a maximum atomic number of only $Z = 68$. The Canadian physicist John McLennan arrived at the same conclusion [McLennan 1923]. Sommerfeld too recognized the problem, but without finding it very serious. He suggested that if the perturbations of the other electrons were taken into account the limit might possibly be raised from 68 to 92, which he found would be "attractive" since it provided an explanation of uranium being the heaviest element.

The question of the number of chemical elements was reconsidered by Walther Kossel in 1928, still on the basis of the old quantum theory. Pointing out the inadequacy of the Bohr-Sommerfeld treatment, he argued that at very small distances modifications of the Coulomb law of force had to be taken into account [Kossel 1928]. While the electrostatic repulsion of two electrons varies as $r^{-2}$, he assumed an additional magnetic attraction proportional to $r^{-4}$. In this case, if the diameter of the innermost K-orbit ($n = 1$) becomes very small, an electron in such a state might fall into the nucleus and reduce its charge. Recall that until the early 1930s it was universally believed that a nucleus characterized by the integers A and Z consisted of A protons and A – Z electrons.

When Sommerfeld's relativistic extension of Bohr's atomic theory was replaced by the quantum-mechanical theory built on the Dirac equation, the energy expression for the lowest bound state in a one-electron system remained unchanged, although the permitted values and the meaning of the quantum numbers were now somewhat different. The first physicist to provide an exact solution, Walter Gordon at the University of Hamburg, commented in a footnote on the problem of a highest atomic number [Gordon 1928]. As a mathematical requirement for solving the Dirac equation for a nuclear charge $Ze$, he found

$$\sqrt{1 - \alpha^2 Z^2} > \tfrac{1}{2},$$

and thus

$$Z < \frac{\sqrt{3}}{2}\frac{1}{\alpha} = 118.7 \pm 0.1$$

This result, he was pleased to note, "is satisfied in the case of the periodic system." In Gordon's treatment, screening corrections due to the presence of other electrons were not taken into account. In general, also with the Dirac theory the lowest permitted energy goes towards zero when $Z$ approaches $1/\alpha$ from below, and it becomes imaginary when $Z > 1/\alpha \cong 137$. A point nucleus with $Z > 137$ cannot support the lowest bound electron.

**4  The minimum-time hypothesis**

In the late 1920s there appeared several ideas of a smallest time interval, that is, a fixed minimum duration $\Delta T$ below which time measuring would have no meaning [Kragh and Carazza 1994]. The minimum time interval, sometimes called a "chronon," was usually assumed to be given by $\Delta T = h/mc^2$, where $m$ is the mass of either an electron or a proton. If a duration cannot be shorter than $\Delta T$, either about $10^{-20}$ s or $10^{-23}$ s, the period and velocity of an atomic K-electron must be similarly limited. This places a limit on the atomic number, such as can be seen from the relationships of the simple Bohr theory, where

$$v = Zc\alpha \quad \text{and} \quad r = h/2\pi mv$$

With $\Delta T = h/mc^2$ and $m$ denoting the electron mass, the condition that the period of revolution must exceed the minimal time limit implies

$$\frac{2\pi}{v} = \frac{h}{mc^2},$$

from which Z < 137. The same result follows, even more simply and without making explicit use of the $\Delta T$ hypothesis, from $v = Zc\alpha$ and $v < c$.

In a paper of 1928, the British physicists Henry Flint and Owen Richardson argued from quantum mechanics and special relativity that

$$\Delta T = h/m_0 c^2 \cong 10^{-20} \text{s}$$

was a minimum proper time unit. The period of revolution, measured in the electron's proper time, must be larger than the postulated time unit:

$$\frac{2\pi r}{v}\sqrt{1-\frac{v^2}{c^2}} > \frac{h}{m_0 c^2}$$

Introducing in this inequality $r = h/2\pi m v$ with $m$ expressed relativistically by $m_0$ leads to

$$1 - \frac{v^2}{c^2} > \frac{v^2}{c^2} \quad \text{or} \quad v < \frac{c}{\sqrt{2}}$$

That is, Flint and Richardson claimed that the velocity of an orbiting K-electron could not exceed 71% of the velocity of light. It then follows immediately from $v = Zc\alpha$ that

$$Z < \frac{1}{\alpha\sqrt{2}} \quad \text{or} \quad Z < 97$$

The two physicists observed that their result did not really refer to the nuclear charge as such, but to the number of electrons in an atom: "The limit is on the charge of a nucleus which can build up a chemical atom. So far as the restriction goes very hot stars might contain nuclei with higher values of N [Z] than those possessed by any chemical element" [Flint and Richardson 1928, p. 641]. As we shall see in the following section, at the time there were several speculations of stellar elements of very high atomic number. Flint returned to the issue a couple of years later, when he repeated that the minimum-time principle had demonstrated a definite limit to the number of existing elements [Flint 1932].

Like Gordon in 1928 had used the Dirac equation to refine the old Bohr-Sommerfeld result, so the German physicists Walter Glaser and Kurt Sitte applied Dirac's theory combined with the Flint-Richardson minimum-time hypothesis [Glaser and Sitte 1934]. In Dirac's theory there is no definite distance or velocity of the electron, but there are quantum-mechanical analogies relating to the average values $r^{-2}$ and $dx_i/dt$. With these analogies Glaser and Sitte found that Bohr's relation $v = Zc\alpha$ remained valid. For the average distance they derived

$$r = \frac{a_0}{Z\sqrt{2}}\left[2(1 - Z^2\alpha^2) - \sqrt{1 - Z^2\alpha^2}\right]^{1/2}$$

Using the criterion that the period of revolution $2\pi r/v$ has to exceed $h/mc^2$, they found the maximum atomic number to be

$$Z_{max} = 90.5 \pm 0.5$$

Given the existence of uranium with Z = 92 the number comes out too small, but Glaser and Sitte argued that the effects caused by the second K-electron would result in a correction that might increase the number to 92. In a footnote they acknowledged a discussion with their colleague at the German Charles University in Prague, the physicist and philosopher Philipp Frank, who had pointed out that the question of a highest atomic number could also be considered from the perspective of Louis de Broglie's old idea of matter waves. One might require the de Broglie wavelength *h/mv* for a bound electron to be greater than the Compton wavelength *h/mc,* meaning that

$$\frac{h}{m_0 v}\sqrt{1-\frac{v^2}{c^2}} > \frac{h}{m_0 c}$$

The inequality leads to the same result as obtained by Flint and Richardson, namely, $v \leq c/\sqrt{2}$ and therefore Z < 97.

Yet another attempt to calculate the maximum atomic number by means of an off-mainstream physical theory was made by the Indian mathematician Vishnu Narlikar (the father of the cosmologist Jayant Narlikar), who in 1932 applied Eddington's so-called *E*-algebra to the problem. According to Eddington, the magic number 137 represented the number of degrees of freedom of a two-particle system. Assuming a one-to-one correspondence between degrees of freedom and independent wave functions, by means of Pauli's exclusion principle this may be interpreted as implying that the maximum number of electrons in an atom is 137, such as also suggested by the Bohr-Sommerfeld argument. Narlikar may have felt that this was an unrealistically large atomic number. At any rate, he modified Eddington's analysis in a way that reduced the number 137 to 92, and from this he concluded that "there can be no element beyond uranium" [Narlikar 1932].

## 5  Cosmic speculations

Superheavy radioactive elements of a hypothetical and unspecified nature played some role in early attempts to understand astrophysical and cosmological phenomena, including the new and mysterious cosmic rays [Kragh 2007]. By 1910 it was known that most of the radioactive elements were descendants of the long-lived elements uranium and thorium, and also that the ratio of uranium to radiogenic lead provided an estimate of the age of the earth of at least one billion years. But where did the uranium come from? How does it come that uranium is still present on the earth and elsewhere in the universe?

If the universe had existed in an eternity of time, such as was generally assumed before World War II, even the most long-lived elements must have transformed into stable elements. One answer might be that uranium and thorium were themselves decay products of even heavier elements. In a lecture of 1911 Arthur Erich Haas, a physicist at the University of Vienna, considered the possibility of "a mother substance of uranium in the form of another and possibly unknown element." However, he found the hypothesis to be absurd. As he pointed out, if uranium was to be explained as a decay product of a still heavier element, then this element would again have to be the product of a still heavier element, and so on *ad infinitum*, ending up with the impossible notion of a primitive mother element of perhaps infinite atomic weight. Haas' alternative was to regard radioactivity as an arrow of time, a decreasing cosmic process that had once had a beginning. "The phenomenon of atomic decay, which probably governs not only radium and uranium but all matter, constitutes an important new objection against the assumption of an eternal world process" [Haas 1912, p. 183].

Haas' argument against radioactive substances heavier than uranium did not prevent physicists from speculating about such hypothetical elements. In his presidential address to the 1923 meeting of the British Association for the Advancement of Science, Rutherford briefly conjectured that the long-lived radioactive elements were the remnants of a much earlier and much more radioactive state of the universe. "It may be," he said, "that the elements, uranium and thorium, represent the sole survivals in the Earth today of types of elements that were common in the long distant ages, when the atoms now composing the Earth were in course of formation" [Rutherford 1923, p. 20]. Following an independent line of thinking, the eminent physical chemist Walther Nernst not only thought that superheavy radioactive elements had once existed, he also thought they were still being formed in the depths of space. Nernst, who received the Nobel Prize in 1920 for his fundamental contributions to chemical thermodynamics, pursued the idea for more than two decades.

A believer in the ether, Nernst argued that it was filled with an enormous amount of electromagnetic zero-point energy, corresponding to an energy density of no less than $1.5 \times 10^{16}$ J/cm$^3$. Out of fluctuations in this energy-rich ether super-radioactive transuranic atoms would be formed, and the energy accompanying their decay would eventually return to the ethereal energy reservoir [Bartel and Huebener 2007, pp. 306-326]. "Strongly radioactive elements are continually being formed from the æther, though naturally not in amounts demonstrable to us," Nernst wrote in 1928. "The sources of the energy of the fixed stars must be looked for in radio-active elements which are of higher atomic weight than uranium" [Nernst 1928, p. 137 and p. 141]. The hypothesis was an essential part not only of Nernst's explanation of stellar energy production, but also of his favoured cosmological view of an eternal steady-state universe.

Nernst thought that the hypothesis of one or more superheavy elements received some support from measurements of the high-energy component of the penetrating cosmic rays. Although admitting its speculative nature, he urged the chemists to "seek by all suitable means this most important element in the earth also" [Nernst 1928, p. 138]. Apparently his call for action was ignored. Seven years later he restated the conjecture of superheavy cosmic elements, now maintatining that it was "in no way particularly hypothetical" [Nernst 1935, p. 520]. The reason for his optimism were the recent reports from Fermi and others concerning artificially produced transuranic elements. In Germany, Lise Meitner and Otto

Hahn were looking for elements heavier than uranium [Sime 1996, pp. 164-169]. Nernst's hypothesis of element formation from the decay of transuranic elements was not accepted by the majority of physicists, who found synthesis of simple atoms a far more natural and attractive hypothesis. For example, this was the view of the British physical chemist S. Bradford Stone, who in a paper of 1930 argued that the elements were formed through the combination of helium and hydrogen nuclei. From considerations of the mass defect in nuclear reactions he was led to conclude an upper limit of about 340 for the atomic weight [Stone 1930].

On the other hand, the German physicist Werner Kolhörster, a pioneer of cosmic-rays physics, found Nernst's speculations of superheavy radioactive elements to be valuable and consonant with his own ideas of the origin and nature of the cosmic rays [Kolhörster 1924]. The physical chemist Paul Günther, a former student of Nernst's, agreed that the hypothesis was "not implausible." He added that one might possibly detect traces of elements with atomic number larger than 92 in the interior of the earth [Günther 1925]. The positive attitude was shared by a few other German scientists. Thus, to the mind of the astronomer Walter Schulze, Nernst's theory was in "complete agreement with the most recent findings" in cosmic-rays studies [Schulze 1930]. He found the idea of superheavy cosmic elements appealing because it offered an explanation of the nature and fluctuations of the cosmic rays. Outside Germany Nernst's hypothesis attracted very little interest.

**6  Jeans' superheavy elements**

While Nernst defended a steady-state universe in dynamic equilibrium, the respected physicist and astronomer James Jeans was convinced that the universe was irreversibly running down, its fate being sealed by the tyranny of the second law of thermodynamics. Yet he shared with Nernst, if for different reasons, the predilection for very heavy radioactive elements in the stars and the nebulae. He also agreed with the German chemist that the universe evolves from the complex to the simple [De Maria and Russo 1990].

In a theory of stellar composition from 1926 Jeans concluded that in the centres of the stars, including the sun, there were elements of "exceptionally high atomic weight," meaning A > 240. "We seem driven," he wrote, "to supposing that the main part, at least, of the sun's energy comes from elements of atomic number higher than 92" [Jeans 1926a, p. 563]. Jeans developed his theory of stellar structure, including the hypothesis of superheavy elements, in his 1928 monograph *Astronomy and Cosmogony* and at other occasions. To put it briefly, the theory resulted in a formula that expressed the ratio $Z^2/A$ for stellar matter by the star's central temperature and some other quantities that could be inferred from observations. From this formula he obtained values for $Z^2/A$ far larger than those of the known elements, corresponding to "atomic weights of thousands at least" [Jeans 1928a: 104]. Realizing that such gigantic atoms were improbable he modified the values appearing in his formula for $Z^2/A$, primarily by reducing the temperature. In this way he was led to atomic numbers in the neighbourhood of $Z = 95$, which he considered to be "entirely consistent with all the known facts" [Jeans 1930, p. 312].

Jeans expected that stars younger and more massive than the sun would consist mainly of the superheavy elements, and that the nebulae would be particularly rich in elements of the highest atomic weights. In the course of time the superheavy elements would transform into radiation, either by proton-electron annihilation or by ordinary radioactive decay. He also suggested the more radical hypothesis that annihilation of entire atoms might occur in the stars. In the case of the sun, he argued that the outer layers were not representative for its chemical composition. The very heavy elements would have sunk to its far interior and thus not be detectable by spectroscopic means. Although the earth was undoubtedly formed by solar matter, according to Jeans it was formed mainly or solely out of the lighter atoms of the sun's surface, and for this reason there would be no traces of the superheavy elements in the crust of the earth.

Admitting that there was not the slightest direct empirical evidence for the superheavy stellar elements, Jeans (like Nernst) justified the hypothesis by what he considered its explanatory power. He believed that without this hypothesis, two important questions would remain unanswered: the nature of stellar energy production and the presence of uranium and thorium in the crust of the earth. With regard to the first problem, he argued that it could not be explained on the basis of the types of matter known to the chemists. "Other types of matter must exist," he said, "and … these other types can only be elements of higher atomic weight than uranium" [Jeans 1926b, p. 37]. Of course he then had to face the question of the origin of the hypothetical superheavy elements. Instead of relying on the energy reservoir of the ether, as Nernst did, he conjectured that matter had not always existed. There had been "a definite event, or series of events, or continuous process, of creation of matter at some time not infinitely remote" [Jeans 1930, pp. 336-337]. At this event matter was created by high-energy photons "being poured into space." Jeans did not explain where the primordial high-energy photons came from. Speaking in the language of metaphors rather than science, he famously proposed that "we may think of the finger of God agitating the ether."

In a lecture given in the autumn of 1928 he repeated that the cores of the stars were rich in transuranic elements. He generalized: "The complete series of chemical elements contains elements of greater atomic weight than uranium, but all have, to the best of our knowledge, vanished from the earth, as uranium is also destined to do in time" [Jeans 1928b, p. 696]. He described terrestrial radioactive elements such as uranium and thorium as "the last surviving vestiges of more vigorous primeval matter, thus forming a bridge between the inert permanent elements and the heavier and shorter-lived elements of the stars" []eans 1928a, p. 135].

Jeans' theory was received no more kindly than Nernst's. It was discussed at a meeting of the Royal Astronomical Society on 11 June 1926 where it was met with strong opposition from Arthur Eddington and Edward Arthur Milne, England's two foremost theoretical astrophysicists. Milne objected that the theory went contrary to the generally held view that the heavy atoms were synthesized in the interior of the stars. This was also the view of Eddington, who on another occasion, alluding to Jeans and Nernst, objected to the assumption "that more potent elements exist beyond uranium, responsible for the larger stellar supply." He considered it contrived as well as anti-evolutionary. "Personally," Eddington said, "when I contemplate the uranium nucleus consisting of an agglomeration of 238 protons and 146 electrons, I want to know how all these have been gathered together" [Eddington 1927-1929, p. 111]. Another response to Jeans' superheavy elements came from the Russian astronomer Boris Gerasimovich and his U.S. colleague Donald Menzel in a joint review article on stellar energy production. The two astronomers dismissed Jeans' postulate as "highly unsatisfactory" and "too highly speculative and artificial to carry much weight" [Gerasimovich and Menzel 1929].[4] As an additional argument against the superheavy elements they referred to the previously mentioned calculations of Bohr, Sommerfeld and Kossel.

---

[4] The theories of Nernst and Jeans were undoubtedly speculative, although in this respect they did not match another theory of superheavy elements proposed in 1926 by Monroe Snyder, a former high school teacher in astronomy [Snyder 1926a; Snyder 1926b]. According to Snyder, the highest atom possible had atomic number 143. The corresponding element, which he named "ultine," was homologuous to chlorine and supposed to play a role in the cosmic rays. Remarkably, Snyder published his amateurish speculations in a distinguished academic journal, the *Proceedings of the American Philosophical Society* founded in 1838. His theory was not taken seriously by the scientists, most of whom were probably unaware of it.

**7  The ultimate superheavy atom**

It is of course possible to conceive of atoms even heavier than the unnamed Bohr-Sommerfeld atom of Z = 137 or Snyder's "ultine" of Z = 143. The truly ultimate limit was reached in 1931, when the Belgian physicist and cosmologist Georges Lemaître proposed the first version ever of big bang cosmology. The existence of radioactive elements with half-lives of the order of $10^9$ years served as an inspiration for his idea of a finite-age exploding universe or what he referred to as the primeval-atom hypothesis. He likened the original compact universe to a huge super-radioactive atomic nucleus with a correspondingly huge atomic number. We could conceive, he said, "the beginning of the universe in the form of a unique quantum, the atomic weight of which is the total mass of the universe." The primeval atom would spontaneously disintegrate, and "Some remnants of this process might, according to Sir James Jeans's idea, foster the heat of the stars until our low atomic number atoms allowed life to be possible" [Lemaître 1931a]. At a conference in London in September 1931 celebrating the centenary of the British Association for the Advanvement of Science Lemaître admitted inspiration from Jeans, who was also present. Indeed, his picture of the primeval atom as one huge atomic nucleus had some similarity to Jeans' superheavy elements, only with the atomic weight extrapolated to the most extreme limit: "Sir James Jeans has given strong reasons for admitting the existence of atoms of considerably higher atomic weight that our actual dead atoms. Cosmogony is atomic physics on a large scale – large scale of space and time – why not large scale of atomic weight?" [Lemaître 1931b, p. 705].

Lemaître did not care to distinguish between the terms "nucleus" and "atom," for the atomic number of the primeval atom and its decay products were so excessively large that it made the distinction illusory. As he pointed out, for elements of very large atomic number, "the K-ring would merge into the nucleus." Although Lemaître did not think of the primeval atom as a chemical element in the ordinary sense, the analogy was part of the imagery that inspired him to propose the big bang hypothesis. At the end of his contribution to the London conference he suggested that to develop what might appear to be a "wild imagination" into a proper physical hypothesis one needed "a theory of atomic structure sufficient to be applied to atoms of extreme weights."

Lemaître's primeval-atom hypothesis was either ignored or rejected as a wildly speculative *jeu d'esprit*. According to the Canadian astronomer John Stanley Plaskett [1933, p. 252] it was "the wildest speculation of all," nothing less than "an example of speculation run mad without a shred of evidence to support it." Among the few who found the hypothesis appealing was the American astronomer Paul W. Merrill, of the Mount Wilson Observatory. In a brief paper on "Cosmic Chemistry" of 1933 he called attention to Lemaître's unusual explanation of the lighter elements as descendants of much heavier elements, a feature it shared with the theories of Nernst and Jeans. "Perhaps," Merrill said, "we are already too late for some of the original heavier elements, but just in time for uranium, thorium, and radium which will, in turn, soon be exhausted. Future chemists may speculate about them just as we speculate about elements heavier than uranium. … Carried to its logical limit the theory postulates an original universe in the form of one immense super-radioactive cosmic atom. It is a daring speculation, but a beautiful and a suggestive one" [Merrill 1933, p. 28].

As it turned out, the transformation of Lemaître's primeval-atom hypothesis into a physical big bang theory did rely on progress in nuclear physics, but not of the kind he had in mind. Yet it is of interest to note that George Gamow, who was chiefly responsible for the transformation, at one occasion suggested that the primeval superdense nuclear matter might consist of superheavy nuclei [Gamow 1942]. He speculated that these hypothetical nuclei – "several times heavier than uranium" – would undergo multiple fission processes. Gamow soon realized that the hypothesis of primordial superheavy elements was a dead end and that big bang cosmology had to start with very simple rather than very complex nuclear particles. He chose neutrons.

**8 Conclusion**

As shown by this review, even before the proton-neutron model of the atomic nucleus – or even before the nuclear atom – several chemists and physicists expressed an interest in the possibility of transuranic elements. Apart from a single discovery claim of 1904 and a few later suggestions based on inconclusive evidence, until the late 1930s the standard view remained that $Z = 92$ is the highest atomic number. If this were indeed the case, the number ought to be explainable in terms of atomic and quantum theory. In the 1920s there were several attempts to establish an upper limit of the periodic system, resulting in either $Z = 92$, $Z = 137$ or $Z = 118$. The physicists doing work along this line did not really believe in the existence of transuranic elements. Realizing the uncertainty of their calculations, they had no problem with accepting uranium as the heaviest of the actually existing elements.

    On the other hand, some physicists and chemists believed that unsolved problems in astrophysics, such as the energy generation of the stars and the enigmatic cosmic rays, required the hypothesis of celestial superheavy elements. This idea was championed by Nernst and Jeans in particular, but it was considered unorthodox and unsatisfactory by the large majority of physicists and

astronomers. Although forgotten today, the speculations about stellar superheavy elements are likely to have acted as inspiration for Lemaître in his revolutionary proposal of an exploding universe. From a modern point of view, what is perhaps the most striking in the development here reviewed is the willingness of scientists to engage in speculations almost completely divorced from empirical data. In stark contrast to the earlier speculations, the development in the 1930s that led to the discovery of the first transuranic elements was experimental and firmly based in the new nuclear physics. It seems to have owed little or nothing to the earlier speculative tradition.